\documentclass[reqno]{amsart}

\usepackage{dsfont}
\usepackage{amscd, graphicx}

\newtheorem{remark}{Remark}

\def\Z{\mathbb{Z}}
\def\R{\mathbb{R}}
\def\C{\mathbb{C}}
\def\H{\mathcal{H}}

\def\a{\alpha}

\def\s{\sigma}
\def\p{\psi}

\def\2{\frac{1}{2}}
\def\ii{\frac{1}{2i}}
\def\Tr{\mathrm{Tr}}

\def\be{\begin{equation}}
\def\ee{\end{equation}}
\def\bp{\begin{proof}}
\def\ep{\end{proof}}
\def\bc{\begin{cases}}
\def\ec{\end{cases}}

\newcommand{\bra}[1]{\ensuremath{\left\langle #1\right|}}
\newcommand{\ket}[1]{\ensuremath{\left|#1\right\rangle}}
\newcommand{\braket}[2]{\ensuremath{\left\langle #1\vphantom{#2}\right.\left|\vphantom{#1}#2\right\rangle}}
\newcommand{\tbraket}[2]{\ensuremath{\left\langle #1\vphantom{#2}\right.\left.\otimes\,\,\vphantom{#1}#2\right\rangle}}

\numberwithin{equation}{section}

\newtheorem{Pro}{Proposition}

\begin{document}

\markboth{P. Aniello, J. Clemente-Gallardo, G. Marmo and G. F. Volkert}
{Classical Tensors and Quantum Entanglement I: Pure States}

%

%

\title{CLASSICAL TENSORS AND\ QUANTUM\ ENTAGLEMENT I: PURE STATES}

\author{P. ANIELLO}

\address{Dip. Sc. Fisiche, Fac. Sc. Biotecnologiche and INFN-Napoli, Universit\`a Federico II\\
Via Cintia, Napoli, 80126, Italy
}

\author{J. CLEMENTE-GALLARDO}

\address{Institute for Biocomputation and Physics of Complex Systems\\
Corona de Arag\'on 42, 50009 Zaragoza, Spain\\
and\\
Departmento de F\'\i sica Te\'orica \\
Universidad de Zaragoza\\
50009 Zaragoza, Spain 
}

\author{G. MARMO}

\address{Dip. Sc. Fisiche and INFN-Napoli, Universit\`a Federico II\\
Via Cintia, Napoli, 80126, Italy
}

\author{G. F. VOLKERT}
\address{Dip. Sc. Fisiche and INFN-Napoli, Universit\`a Federico II\\
Via Cintia, Napoli, 80126, Italy\\
and\\
Mathematisches Institut, Ludwigs-Maximilians-Universit\"at\\ 
Theresienstr. 39, 80333 M\"unchen, Germany\\
}


\maketitle

\begin{abstract}
The geometrical description of a Hilbert space asociated with a quantum system considers a Hermitian tensor to describe the scalar inner product of
vectors which are now described by vector fields. The real part of this
tensor represents a flat Riemannian metric tensor while the imaginary part
represents a symplectic two-form. The immersion of classical manifolds in the
complex projective space associated with the Hilbert space allows to
pull-back tensor fields related to previous ones, via the immersion
map. This makes available, on these selected manifolds of states, methods of
usual Riemannian and symplectic geometry. Here we consider these pulled-back tensor fields when the immersed submanifold contains separable states or entangled states. Geometrical tensors are shown to encode some properties of these states. These results are not unrelated with criteria already available in the literature. We explicitly deal with some of these relations.
\end{abstract}


\section{Introduction}
In a previous paper \cite{Aniello:08}, we have shown how to associate classical manifolds of quantum states with unitary representations of Lie groups. This procedure resembles what has been done by Peremolov in defining generalized coherent states \cite{Perelomov:1971bd}. Our approach is different because we use groups only in some instrumental manner to identify submanifolds of quantum states  with orbits originated from some fiducial state. We are not intersted in "comparing" classical and quantum evolution as done for coherent states, our aim is simply to use classical methods of differential geometry to describe selected manifolds of states of the quantum system we are considering. Of course the total complex projective space itself can be considered as a
"classical manifold", however due to the present state-of-the art of
infinite dimensional differential geometry, methods from differential
geometry are much more effective when the identified submanifold has finite
dimension. Fortunately many situations of great physical interest like those emerging in quantum computation are concerned with finite dimensional manifolds of quantum states \cite{Horodecki:2007}\cite{Bengtsson:2006}\cite{Brody:1999cw}.\\
In this paper, by using the approach of our previous papers \cite{Aniello:08}\cite{ClementeGallardo:2008ey}\cite{Carinena:2007ws}\cite{Grabowski:2000zk}\cite{Grabowski:2005my}, we are going to consider the information that can be extracted from pulled-back tensor fields when we consider states of composite systems acted upon by local unitary groups, the so called gauge groups which do not change the entanglement properties of the starting state. In some sense these submanifolds are made of quantum states with the "same content of entanglement". Clearly the procedure can handle bipartite systems and multipartite systems. In this paper however we shall limit ourselves to
bipartite systems. How it will be clear from the text, the selected
submanifold of quantum states need not be associated with unitary
representations of Lie groups, as a matter of fact within the stratified
manifold of states, strata may be identified with orbits of the general
linear group while states on the orbits of the unitary local group not only
have the same rank of the initial quadratic form but also the same
"spectrum"\cite{Grabowski:2000zk}\cite{Grabowski:2005my}. We are confident that methods from differential geometry will turn out to be useful whenever some spaces of interest (like for instance spaces of entangled
states) do not carry a linear structure.\\
We shall review very briefly how covariant tensor fields may be pulled-back from the complex projective space to the identified submanifold of
states. Then we consider the particular situation of composite systems and
analyze the pulled-back tensor when the starting state is separable and when
the starting state is entangled. Thereafter we compare the entanglement as
detected by our procedure with other measures of entanglement available in
the literature, with particular attention to those which seem to be very similar.

\section{Tensors on Orbits of Unitary related Quantum States}\label{Tensors on orbits of unitary related quantum states}

Let us start by considering a finite dimensional Hilbert space $\H=\C^k$ in a
basis $\{e_j\}_{j\in J}$ with a corresponding family of coordinate functions  
\be \braket{e_j}{\p}=z^{j}(\p)=z^{j}.\ee 
By replacing functions with their exterior differential, we may associate with
\be \braket{\p}{\p}=\sum_{j}\bar{z}^{j}z^{j},\ee
a Hermitian covariant tensor
\be\tbraket{d\p}{d\p}:= \sum_{j} d\bar{z}^{j}\otimes dz^{j}.\label{geom}\ee
With vectors $u$ and $v$ we associate vector fields $X_u:\p\mapsto (\p, u)$ and
$X_v:\p\mapsto (\p, v)$ and we have \cite{ClementeGallardo:2008ey} 
 \be\tbraket{dz}{dz}(X_w,X_v)=\braket{w}{v}.\ee
The decomposition of the coordinate functions $\braket{e_j}{\p}=z^{j}$  into
real and imaginary part 
\be z^{j}=q^{j}+ip^{j}\ee
decompose the Hermitian tensor $\tbraket{dz}{dz}$
into
\be\sum_{j} dq^{j}\otimes dq^{j}+dp^{j}\otimes dp^{j}+i(dp^{j}\otimes dq^{j}-dp^{j}\otimes dp^{j}),\ee
i.e.\,an Euclidean and a symplectic tensor on $\H_{\R}:=\R^{2n}$.\\
Consider next a general manifold $Q$ with $\text{Dim}_{\R}(Q)\subset \H_{\R}$ and an embedding 
\be\iota_Q: Q \hookrightarrow \H.\ee
A covariant Riemannian metric tensor $G$ on $Q$ can then be defined by the real part of the induced pull-back $\iota^*_Q$ of the Hermitian tensor $\tbraket{d\p}{d\p}$, where else its imaginary part yields a closed two-form $\Omega$, which becomes in a non-degenerate case, symplectic \cite{Aniello:08}. 
In short,
\be G:=\text{Re}(\iota^*_Q(\tbraket{d\p}{d\p})).\ee
\be \Omega:=\text{Im}(\iota^*_Q(\tbraket{d\p}{d\p})),\ee
where we adopt the notation for the pulled back Hermitian tensor
\be \iota^*_Q(\tbraket{d\p}{d\p})=  G + i \Omega.\ee
When the manifold $Q$ admits the structure of a Lie-group, we may identify a submanifold of states by choosing a normalized fiducial state
\be\ket{0}\in \H\label{emb1}\ee
on which we act with $q\in Q$:
\be U(q)\ket{0}\in \H\ee
via a unitary representation $U(q)$ of the preassigned Lie group $Q$. The structure of the orbits, which are generated by the unitary group actions, will depend on the choice of the fiducial state $\ket{0}$. Each orbit may be identified with the quotient space
\be Q/Q_{\ket{0}}\cong\{U(q)\ket{0}: q\in Q\}/\sim,\label{emb2}\ee
where $Q_{\ket{0}}$ is the stabilizer or isotropy group of $\ket{0}$. 
Under regularity conditions this orbit becomes then identified with a submanifold of the Hilbert space, similar to the case of Perelomov's generalized coherent states \cite{Perelomov:1971bd}.\\
With a given pair $(U(q),\ket{0})$, a unitary representation and a fiducial state, we associate the embedding action
\be  U(q)\ket{0} \equiv \ket{q}.\label{emb-action}\ee
The unitary representation of the Lie group defines a representation $R$ of its Lie-algebra. It is defined by means of the tangent map of the representation map. We have
\be R: T_e Q \to T_e U(\H),\label{map to operators}\ee
\be  R([e_j, e_k])=[R(e_j), R(e_k)].\ee
The exterior derivative on states (\ref{emb-action}), considered as vector-valued functions on the group orbit
gives
\be \ket{dq}=dU(q)\ket{0}=UU^{-1}dU(q)\ket{0},\ee
i.e.\,we may also define operator valued 1-forms
\be U(q)^{-1}dU(q)\equiv iR(e_j)\theta^j\label{1-forms}\ee
and a specific expression for the differential of the group action (representation)
\be dU(q)=iU(q)R(e_j)\theta^j.\label{decomposition}\ee
It relies on:
\begin{itemize}
\item[1.] the Lie-group action $U(q)$
\item[2.] a set of Hermitian operators $R(e_j)$, and 
\item[3.] a family $\{\theta^j\}_{j\in J}$ of left-invariant 1-forms.
\end{itemize}
In conclusion, (\ref{map to operators}) defines a map from $T_eQ$ to essentially self-adjoint operators acting on $\H$. With any state \ket{\p} we associate a fiber-wise linear function on TQ by means of 
\be\frac{\bra{\p}R(e_j)\ket{\p}}{\braket{\p}{\p}}\theta^j.\ee

If we introduce a tensor-product \be \otimes:=\otimes_{\mathcal{A}}\ee on the Algebra
$\mathcal{A}$ of 'classical' 1-forms spanned by $\{\theta^j\}_{j\in J}$, we find by
using the decomposition (\ref{decomposition}), an  \emph{operator}-valued
(0,2)-tensor field: 
$$ dU(q)^{\dagger}\otimes dU(q)$$
$$=R(e_j)^{\dagger}U(q)^{\dagger}\theta^j \otimes U(q)R(e_k)\theta^j$$
$$ = R(e_j)^{\dagger}U(q)^{\dagger}U(q)R(e_k)\theta^j \otimes \theta^k$$
\be = R(e_j)R(e_k)\theta^j \otimes\theta^k.\ee
Expectation values of this operator-valued tensor on the chosen fiducial state  
\be \bra{0}R(e_j)R(e_k)\ket{0}\theta^j \otimes\theta^k.\label{PBT}\ee
provide 'classical' tensors on the manifold of quantum states obtained from
$\ket{0}$ with the action of the Lie group $Q$, yielding a symmetric
part\footnote{We adopt here the following shorthand notations: $$\theta^j
  \odot\theta^k:=\2(\theta^j \otimes\theta^k+\theta^k \otimes\theta^j)$$ 
$$\theta^j \land\theta^k:=\2(\theta^j \otimes\theta^k-\theta^k \otimes\theta^j)$$
$$[R(e_j),R(e_k)]_+:=\2(R(e_j)R(e_k)+R(e_k)R(e_j))$$
$$[R(e_j),R(e_k)]_-:=\ii(R(e_j)R(e_k)-R(e_k)R(e_j)).$$\label{FT}}
\be \bra{0}[R(e_j),R(e_k)]_+\ket{0}\theta^j \odot\theta^k,\ee
which defines a metric tensor, and an antisymmetric part
\be i\bra{0}[R(e_j),R(e_k)]_-\ket{0}\theta^j \land\theta^k,\label{APBT}\ee
yielding a closed 2-form on $Q/Q_{\ket{0}}$. The latter becomes a symplectic
tensor, whenever it turns out to have a trivial kernel.\\ 
In general we may use any state, say a positive normalized functional $\rho \in
u^*(\H)$ and consider, in analogy with (\ref{PBT}), the classical tensor on the
group manifold associated with the quantum density state $\rho$ 
\be \rho (R(e_j)R(e_k))\theta^j \otimes\theta^k,\ee
being again expanded in a real symmetric part and an imaginary skew-symmetric part
\be \rho ([R(e_j),R(e_k)]_+)\theta^j \odot\theta^k,\ee
and
\be \rho (i[R(e_j),R(e_k)]_-)\theta^j \land\theta^k,\ee 
respectively.
\begin{remark}
Starting from the operator-valued (0,2)-tensor field $dU^{\dagger}(q)\otimes dU(q)$ we
could define directly  
\be \rho(dU^{\dagger}(q)\otimes dU(q)),\ee
amounting a left invariant tensor field on the group manifold. Similarly to
what happens in the GNS construction, this tensor will not be
not-degenerate. It will be degenerate along the intersection of $R(T_eQ)$ with
the Gelfand ideal associated with $\rho$. Therefore the tensor is not degenerate
on the quotient space $Q/Q_{\rho}$,  $Q_{\rho}$ being the group associated with the
algebra in the Gelfand ideal.   
\end{remark}

It is important to notice at this point that the way we imbed our group in the
carrier space we are interested in depends on the action we choose. For
instance we may consider the co-adjoint action on the space of states,
identified with the affine subspace of $u^*(\H)$ defined by $\Tr(\rho)=1$. 
We have 
\be \rho = U(q)\rho_0 U(q)^{\dagger}\label{unitary orbits}\ee
and define the differential $d \rho$ given by
\be d \rho(q) = [dU(q)U^{-1}(q), \rho_0]\ee  
which one obtains by direct computation.

  \begin{remark}
We should briefly explain the meaning of exterior derivative on the stratified
manifold of states. Whenever we consider $d\rho$, when $\rho$ is a state, the
differential should be understood as taken on $u^*(\H)$, i.e. the differential
calculus on the stratified manifold is the one inherited from the ''ambient
space'' $u^*(\H)$.     
  \end{remark}

Again we find an operator-valued $(0,2)$-tensor 
\be d\rho \otimes d\rho\ee 
which may be turned into a (0,2)-tensor on the group manifold by taking the evaluation on the state $\rho$ itself
\be \Tr(\rho d\rho\otimes d\rho).\ee  
It is not difficult to see that this tensor is equal to 
$$ \Tr(\rho (U[iR(e_j),\rho_0]U^{\dagger}U [iR(e_k),\rho_0] U^{\dagger}))\theta^j \otimes\theta^k$$
\be =-\Tr(\rho_0 [R(e_j),\rho_0][R(e_k),\rho_0] )\theta^j \otimes\theta^k.\ee
The structure of this tensor shows clearly that it is degenerate along the
commutant of $\rho_0$, therefore it is not degenerate on the homogeneous space
$Q/Q_{\rho_0}$.\\ 
By focusing on the coefficients of the tensor one finds that they decompose in
a sum of three terms    
\be \Tr(\rho_0^3 R(e_j)R(e_k)) - 2\Tr(\rho_0^2 R(e_j)\rho_0R(e_k)) + \Tr(\rho_0^2
R(e_k)\rho_0R(e_j)).\ee   
These terms reduce in case of pure states, i. e. for $\rho_0^2=\rho_0$ to 
\be \Tr(\rho_0 R(e_j)R(e_k)) - \Tr(\rho_0 R(e_j)\rho_0R(e_k)),\ee 
corresponding to the coefficients of a tensor
\be K:=\Big(\text{Tr}(\rho_0 R(e_j)R(e_k))-\text{Tr}(\rho_0 R(e_j))\text{Tr}(\rho_0
R(e_k))\Big)\theta^j \otimes\theta^k.\label{ProjectivePBT}\ee 
Let us underline that this tensor $K$ will be from now on fundamental for our
following considerations. To see explicitly what this tensor yields when the
density state is a pure state we consider \be \rho_{\ket{0}}\equiv
\frac{\ket{0}\bra{0}}{\braket{0}{0}},\ee 
we find  
\be
\Bigg(\frac{\bra{0}R(e_j)R(e_k)\ket{0}}{\braket{0}{0}}-
\frac{\bra{0}R(e_j)\ket{0}\bra{0}R(e_k)\ket{0}}{\braket{0}{0}^2}\Bigg)\theta^j 
\otimes\theta^k,\label{tbp2}\ee 
a tensor on the punctured Hilbert space $\H_0:=\H-\{0\}$,
whenever we restrict $\rho_0$, to be defined as a pure state associated to the
fiducial state $\ket{0}\in \H$. 

  \begin{remark}
This tensor coincides with a Hermitian tensor 
\be \frac{\tbraket{d\p}{d\p}}{\braket{\p}{\p}}-\frac{\braket{\p}{d\p}\otimes \braket{d\p}{\p}}{\braket{\p}{\p}^2},\ee
on $\H_0$, when pulled back on the corresponding orbit being embedded in $\H$
as described in (\ref{emb1})-(\ref{emb2}). It is invariant under the
multiplicative action of $\C_0$.
  \end{remark}

We shall denote by $\rho_0$ what should be written as $\rho_{\ket{0}}$ and adopt the
notation for the the complex valued tensor coefficients 
\be K_{jk}= \text{Tr}(\rho_0 R(e_j)R(e_k))-\text{Tr}(\rho_0 R(e_j))\text{Tr}(\rho_0 R(e_k)),\ee
admitting a decomposition 
\be K_{jk}=K_{(jk)}+iK_{[jk]} \ee
into a real and symmetric 
\be K_{(jk)}= \text{Tr}(\rho_0 [R(e_j),R(e_k)]_{+})-\text{Tr}(\rho_0
R(e_j))\text{Tr}(\rho_0 R(e_k)),\label{SEV}\ee 
and an imaginary and anti-symmetric part
\be iK_{[jk]}= i\text{Tr}(\rho_0 [R(e_j),R(e_k)]_{-}).\label{ASEV}\ee
We finally remark that the construction of the pull-back tensor depends on the
choice of the fiducial quantum state $\rho_0 \in u^*(\H)$, and on the choice of the
representations of the Lie group along with the associated Lie algebra
representation. On the other hand we recall that left invariant 1-forms 
\be \theta^j:Q\to T^*Q \cong Q \times T_e^*Q\ee 
provide a trivialization of the cotangent bundle of the Lie group and do not
depend neither on the chosen representation nor on the fiducial state. To
discuss the dependence of the pulled-back tensor  
\be K= K_{jk}\theta^j\otimes\theta^k \ee
on the choice of fiducial quantum state $\rho_0$, it may therefore turn out
sufficient to focus only on the coefficients $K_{jk}$ of the tensor.

\section{Orbits of Local Unitary Transformations on Entangled Bi-partite States}\label{Local unitary orbits of entangled bi-partite states}
\subsection{Qualitative statements}

Let us apply the pull-back procedure to orbits 
of quantum states of a bi-partite composite system 
\be \H = \H_1\otimes \H_2\ee
with  
\be  \H_1 \cong \H_2 \cong \C^N.\ee
The orbit will be defined by a family of \emph{local} unitary transformations, realized in terms of a product representation
\be  Q\equiv U(N)\times U(N) \to \text{Aut}(\H_1 \otimes \H_2 )\ee
\be  q\equiv(g, g') \mapsto U(q)\equiv U_1(g)\otimes U_2(g').\ee
The later may be decomposed into
\be  U_1(g)\otimes U_2(g')=(U_1(g)\otimes \mathds{1}) \cdot (\mathds{1}\otimes U_2(g')) .\label{U2U2}\ee
The corresponding infinitesimal action of each factor
$U_1(g)\otimes \mathds{1}$, resp. $\mathds{1}\otimes U_2(g')$ may be realized by means of $N^2-1$ Hermitian matrices which are a generalization of Pauli-matrices (i.e. they are traceless and trace-orthonormal)
\be \sigma_a \in T_eU(N),\ee 
with $1 \leq a \leq N^2-1$ and $\sigma_0 := \mathds{1}$, accordingly we define
\be R(e_j) =
 \begin{cases}
 \sigma_j\otimes \mathds{1} &\text{ for } 1 \leq j \leq N^2\\
 \mathds{1}\otimes  \sigma_{j-N^2} &\text{ for } N^2+1 \leq j \leq 2N^2, 
\end{cases}\label{h-basis}
 \ee
each of them may be considered as the infinitesimal generators of one-dimensional subgroup of the real parameterized Lie-group $U(N)\times U(N)$ after multiplication by the imaginary unit $i$. This is the basis we are going to use in the following, where for both subsystem  bases $\{i\sigma_a\}_{a\in J}$ we adopt the short hand description of indices $a,b$ with $1 \leq a, b \leq N^2$ (instead of using $N^2+1 \leq j-N^2, k-N^2 \leq N^2$ for the second subsystem). By applying the matrix product between two realizations in (\ref{h-basis}), we will get three distinguishable classes of combinations, each yielding non-Hermitian matrices of the form
\be R(e_j)R(e_k) =
 \begin{cases}
 (\sigma_a\otimes \mathds{1})(\sigma_b\otimes \mathds{1})=\sigma_a\sigma_b\otimes \mathds{1}\\
 (\mathds{1}\otimes  \sigma_a)(\mathds{1}\otimes \sigma_b)=\mathds{1}\otimes \sigma_a\sigma_b\\
(\sigma_a\otimes \mathds{1})(\mathds{1}\otimes  \sigma_b)=\sigma_a\otimes \sigma_b.
\label{Product}\end{cases}
\ee
Let us consider the (anti-) symmetrization of (\ref{Product}), 
\be [R(e_j),R(e_k)]_{\pm} =\begin{cases}
 [\sigma_a,\sigma_b]_{\pm}\otimes \mathds{1}\\
\mathds{1}\otimes[\sigma_a,\sigma_b]_{\pm}\\
 \frac{1}{2(i)_-}(\sigma_a\otimes\sigma_b \pm \sigma_a\otimes\sigma_b),
\end{cases}\ee
where in the last line $(i)_-$ denotes the imaginary $i$ to be applied only for the anti-symmetric case. 
By using the commutation relations\footnote{We use here the same (anti)-commutation notation as in footnote \ref{FT}.} 
\be [\sigma_a,\sigma_b]_- = \epsilon_{abc}\sigma_c\ee
and the anti-commutation relations
\be [\sigma_a,\sigma_b]_+ =\frac{2}{N}\delta_{ab}\sigma_0+d_{abc} \sigma_c.\ee
of the generalized Pauli-matrices, we find here the anti-Hermitian matrices
\be [R(e_j),R(e_k)]_- =\begin{cases}
  \epsilon_{abc}\sigma_c\otimes \mathds{1}\\
\mathds{1}\otimes \epsilon_{abc}\sigma_c\\
0
\end{cases}\label{NN-}\ee
for the first two non-trivial cases, and the Hermitian matrices
\be [R(e_j),R(e_k)]_+ =\begin{cases}
\frac{2}{N}\delta_{ab}\sigma_0\otimes \mathds{1}+d_{abc} \sigma_c\otimes \mathds{1}\\
\mathds{1}\otimes \frac{2}{N}\delta_{ab}\sigma_0+\mathds{1}\otimes d_{abc}\sigma_c\\
\sigma_a\otimes\sigma_b. 
\end{cases}\label{NN+}\ee 
Let us consider a given fiducial, pure state 
\be \rho_0 \in \mathcal{P}(\H_1\otimes H_2),\ee
to derive the Hermitian pulled-back tensor $K$ given by (\ref{ProjectivePBT}), on the orbit  
\be \mathcal{O}_{\rho_0}:= U(N)\times U(N)/Q_{\rho_0}\ee 
with isotropy group $Q_{\rho_0}$. We use the above commutator and anti-commutator relations (\ref{NN-}), (\ref{NN+}) to find 
$$ K_{[jk]}:= \text{Tr}(\rho_0 [R(e_j),R(e_k)]_{-}),$$
$$ K_{(jk)}:= \text{Tr}(\rho_0 [R(e_j),R(e_k)]_{+})-\text{Tr}(\rho_0 R(e_j))\text{Tr}(\rho_0 R(e_k)),$$
which have been derived in the previous section in (\ref{SEV}), (\ref{ASEV}). Here we end up with the anti-symmetric components
\be K_{[jk]}= \begin{cases}
\text{Tr}(\rho_0  \epsilon_{abc}\sigma_c\otimes \mathds{1})\\
\text{Tr}(\rho_0\mathds{1}\otimes \epsilon_{abc}\sigma_c)\\
0,
\end{cases}\label{KK-}\ee
where else for the symmetric components we get 
\be K_{(jk)}= \begin{cases}
\frac{2}{N}\text{Tr}(\rho_0(\delta_{ab}\sigma_0\otimes \mathds{1}+d_{abc} \sigma_c\otimes \mathds{1})) -\text{Tr}(\rho_0\sigma_a\otimes \mathds{1})\text{Tr}(\rho_0\sigma_b\otimes \mathds{1})\\
\frac{2}{N}\text{Tr}(\rho_0(\mathds{1}\otimes \delta_{ab}\sigma_0+\mathds{1}\otimes d_{abc}\sigma_c))-\text{Tr}(\rho_0\mathds{1}\otimes  \sigma_a)\text{Tr}(\rho_0\mathds{1}\otimes \sigma_b)\\
\text{Tr}(\rho_0\sigma_a\otimes\sigma_b)-\text{Tr}(\rho_0 \sigma_a\otimes \mathds{1})\text{Tr}(\rho_0\mathds{1}\otimes  \sigma_b).
\end{cases}\label{KK+}\ee
Using these relations we derive the main statement of the present paper:
\begin{Pro}
Let $\mathcal{O}_{\rho_0}$ be an orbit of quantum states related, by means of local unitary transformations, to a pure quantum state $\rho_0\in \mathcal{P}(\C^N\otimes\C^N)$. The pulled back Hermitian tensor $K$, defined in (\ref{ProjectivePBT}), with coefficients
\be K_{jk}= \Tr(\rho_0 R(e_j)R(e_k))-\Tr(\rho_0 R(e_j))\Tr(\rho_0 R(e_k)),\ee
will give rise to
\begin{itemize}
\item[(a)]  vanishing symplectic tensor coefficients $K_{[jk]}$, if $\rho_0$ is
  maximal entangled; 
\item[(b)] a direct sum decomposition $K^1\oplus K^2$ into two Hermitian tensors
  $K^1$, $K^2$, iff $\rho_0$ is separable. 
\end{itemize}
\end{Pro}
\bp
(a) For the anti-symmetric coefficients in (\ref{KK-}) we find  
\be K_{[jk]}= \begin{cases}
\text{Tr}(\text{Tr}_1(\rho_0)  \epsilon_{abc}\sigma_c)\\
\text{Tr}(\text{Tr}_2(\rho_0) \epsilon_{abc}\sigma_c)\\
0,
\end{cases}\label{S-mix=0}\ee
where $\text{Tr}_1(\rho_0)$, $\text{Tr}_1(\rho_0)$ denote here the partial traces,
resp. the reduced density matrices. According to the von Neumann entropy
measure of entanglement for pure states given by  
\be  -\text{Tr}(\text{Tr}_2(\rho_0)\ln\text{Tr}_2(\rho_0)),\ee
it follows that a pure state $\rho_0$ will be maximal entangled, iff it is maximal
mixed in the reduced state, yielding the form 
\be \text{Tr}_2(\rho_0)=\frac{1}{N}\mathds{1}.\ee 
Statement (a) follows then from the traceless property of the Hermitian matrices $\epsilon_{abc}\sigma_c$.\\
(b) For the symmetric coefficients in (\ref{KK+}) we find for $1 \leq j \leq N^2$ and
 $N^2+1 \leq k \leq 2N^2$
\be K_{(jk)}= \text{Tr}(\rho_0\sigma_a\otimes\sigma_b)-\text{Tr}(\rho_0 \sigma_a\otimes \mathds{1})\text{Tr}(\rho_0\mathds{1}\otimes  \sigma_b).\label{Sep-cond}\ee
On the other hand one finds for a given $N\times N$ Bi-partite density state
\be \rho\in u^*(\C^N\otimes\C^N)\ee
its corresponding Fano-Form
\be \rho\equiv \frac{1}{N^2}(\lambda_0\s_0\otimes\s_0 +  n_a \s_a\otimes\s_0 +m_b \s_0\otimes\s_b +t_{ab} \s_a\otimes\s_b).\label{Fano-Form}\ee
For pure bi-partite states one has then, according to \cite{de Vicente:2007}\cite{Man'ko:2002ti}, that $\rho$ will be separable iff the relation
\be  t_{ab}-n_a m_b=0\label{t-nm}\ee
holds. Here one notes that the left hand side in (\ref{t-nm}) becomes identical with (\ref{Sep-cond}), whenever we set $\rho_0\equiv\rho$. With this one concludes
\be K_{(jk)}=0 \text{ for } 1 \leq j \leq N^2 \text{ and }
 N^2+1 \leq k \leq 2N^2,\label{R-mix=0}\ee
i.e.\,vanishing (mixed) Riemannian tensor coefficients, iff $\rho_0$ is separable. 
Finally, one finds by applying the cases (\ref{Product}) on a separable pure state $\rho_0\equiv\rho_1\otimes\rho_2$ that the Hermitian tensor coefficients become 
\be K_{jk}= \begin{cases}
\text{Tr}((\rho_1\otimes\rho_2) \sigma_a\sigma_b\otimes \mathds{1} )-\text{Tr}((\rho_1\otimes\rho_2) \sigma_a\otimes \mathds{1} )\text{Tr}((\rho_1\otimes\rho_2) \sigma_b\otimes \mathds{1})\\
\text{Tr}((\rho_1\otimes\rho_2) \mathds{1}\otimes \sigma_a\sigma_b)-\text{Tr}((\rho_1\otimes\rho_2) \mathds{1}\otimes  \sigma_a)\text{Tr}((\rho_1\otimes\rho_2) \mathds{1}\otimes \sigma_b)\\
0,
\end{cases}\ee
where the third equality follows from the vanishing pre-symplectic coefficients in (\ref{S-mix=0}) and the vanishing  Riemannian coefficients in (\ref{R-mix=0}). Therefore one concludes
\be K_{jk}= \begin{cases}
\text{Tr}(\rho_2)(\text{Tr}(\rho_1\sigma_a\sigma_b)-\text{Tr}(\rho_1 \sigma_a)\text{Tr}(\rho_1 \sigma_b)):=K^1_{ab}\\
\text{Tr}(\rho_1)(\text{Tr}(\rho_2\sigma_a\sigma_b)-\text{Tr}(\rho_2 \sigma_a)\text{Tr}(\rho_2 \sigma_b)):=K^2_{ab}\\
0,
\end{cases}\ee
where $K_{ab}^1$, $K_{ab}^2$ yield tensor coefficients related to Hermitian pulled back tensors, each defined on a orbit $\mathcal{O}_{\rho_1}$, resp. $\mathcal{O}_{\rho_2}$. Vice versa, given a direct sum decomposition $K^1\oplus K^2$ into two Hermitian tensors $K^1,K^2$, we find
\be K_{jk}=K_{(jk)}+iK_{[jk]}= 0 \text{ for } 1 \leq j \leq N^2 \text{ and }
 N^2+1 \leq k \leq 2N^2,\label{K-mix=0}\ee
implying (\ref{R-mix=0}), since (\ref{S-mix=0}) is in general valid, independently of the separability of the state $\rho_0$.
\ep
We subsume that the third class of combinations $\sigma_a\otimes \sigma_b$, having its origin in (\ref{Product}), admits a crucial role in connection with non-local correlations, i.e.\,quantum entanglement, whenever we compute the pulled-back Hermitian tensor (\ref{ProjectivePBT}) on orbits of local unitary related pure bi-partite states. The first and the second kind of combinations $\sigma_a\sigma_b\otimes \mathds{1}$, $\mathds{1}\otimes\sigma_a\sigma_b$ in (\ref{Product}) on the other hand, turn out to be associated to the separability of individual subsystem parts of the bi-partite system. One may in particular detect here a vanishing separability in terms of a vanishing symplectic tensor for maximal entangled states, coherent with the the remarks done in \cite{Bengtsson:2001yd} \cite{Bengtsson:2007}.\\
From our geometric point of view, we may conclude that this dichotomy between separability and entanglement is neatly translated in terms of the decomposition
\be K_{jk}=K_{(jk)}+iK_{[jk]} \ee
of the pulled back Hermitian tensor into an anti-symmetric imaginary part and a symmetric real part, yielding two ,classical' tensors, namely a pre-symplectic tensor (encoding the separability parts $\sigma_a\sigma_b\otimes \mathds{1}$, $\mathds{1}\otimes\sigma_a\sigma_b$) and a Riemmanian tensor (encoding the entanglement part $\sigma_a\otimes \sigma_b$). 

\subsection{Quantitative statements}
For the purpose to proceed from qualitative to quantitative statements on
entanglement we may focus in the following on the Riemannian part. The
Riemannian part decomposes in this regard in block-diagonal matrices $A, B$ and
two equal block-off-diagonal matrices $C$, due to 
\be K_{(jk)}=\left(\begin{array}{cc}A & C \\C & B\end{array}\right).\ee
By identifying the left invariant forms $\theta^a$ of the local unitary subgroups
with a subscript (1), resp. (2), we may associate to each block matrix a
corresponding sub-tensorial quantity 
\be K^A :=A_{(ab)}\theta^a_{(1)}\odot \theta^b_{(1)}\ee
\be K^B :=B_{(ab)}\theta^a_{(2)}\odot \theta^b_{(2)} \ee
\be K^C :=C_{(ab)}\theta^a_{(1)}\odot \theta^b_{(2)}. \ee
The coefficients 
\be  A_{(ab)}=\frac{2}{N}\delta_{ab}+ \frac{1}{2N}n_c d_{abc} -n_a n_b\ee 
\be  B_{(ab)}=\frac{2}{N}\delta_{ab}+ \frac{1}{2N}m_c d_{abc} -m_a n_b\ee  
\be C_{(ab)}=t_{ab}-n_a m_b\label{e-tensor},\ee
can be derived by applying the Fano-Form (\ref{Fano-Form}) in (\ref{KK+}).
According to the previous section we underline again that the latter
coefficients associated to the block-off-diagonal matrices $C$, are those
components which are responsible for the entanglement correlations. In this
regard it becomes natural to search for an entanglement measure associated to
this structure coming along the coefficients of $C$. Indeed by studying the
literature \cite{Schlienz:1995}, one finds an entanglement monotone associated
to $C$ in terms of a normalized trace 
\be \frac{N^2}{4(N^2-1)}\text{tr}(C^TC).\label{e-monoton}\ee
Moreover, this entanglement monotone can be directly related to a measure
proposed in \cite{Man'ko:2002ti}, which is constructed by considering the
partial traces $\rho_j:=\Tr_j(\rho)$ of a mixed bi-partite state $\rho$ and the
difference 
\be R:= \rho - \rho_1\otimes\rho_2.\ee
The entanglement measure is then identified by   
\be \Tr(RR^{\dagger})=\Tr(\rho^2)-\Tr(\rho_1^2\otimes \rho_2^2)-2\Tr(\rho\rho_1\otimes\rho_2).\label{RR}\ee
A straight forward computation in the Fano-Form representation (\ref{Fano-Form}) yields
\be \Tr(RR^{\dagger}) =\frac{1}{N^4} \Tr(C^TC).\ee
This allows to give $\Tr(C^TC)$ a geometric interpretation, namely as a
distance between entangled and separable states \cite{Man'ko:2002ti}.\\ 
Another relation between the classical tensor structure proposed here to the
existing literature on quantum entanglement quantification can be seen provided
by the dimensional characterization of local unitarily generated orbits of
quantum states by means of \emph{Gram matrices}, proposed by Kus et. al. in
\cite{Kus:2001}. The Gram matrix coincides with the pulled-back tensor
considered here in the Riemannian part. An interesting observation in this
regard is that the eigenvalues of a Gram matrix are directly related to the
notion of \emph{concurrence} in the particular case of pure $2\times 2$ systems
\cite{Kus:2001}, an entanglement monotone proposed by Wootters
\cite{Wootters:1997id}. The search for entanganlement monotones via Gram
matrices in arbitrary finite dimensional pure bi-partite systems has been
proposed in a subsequent work \cite{Kus:2002}.

\section{Example: Two entangled qubits}

We may illustrate the pull back procedure and its related statements on the simplest, non-trivial example. By considering a normalized fiducial state \be \ket{0}\in \H = \C^2\otimes\C^2 \cong \C^4,\ee  in a Schmidt-decomposition given by
\be \ket{0}:=\ket{0}_{\alpha_0}= \cos(\alpha_0)\begin{pmatrix}
1 \\
0 
\end{pmatrix}\otimes \begin{pmatrix}
1 \\
0
\end{pmatrix} + \sin(\alpha_0)\begin{pmatrix}
0 \\
1 
\end{pmatrix}\otimes \begin{pmatrix}
0 \\
1
\end{pmatrix},\label{0}\ee
we find the associated pure density matrix
$\rho_0 :=\ket{0}\bra{0} \in \C P^3$, 
\be \left(\begin{array}{cccc}\cos^2(\alpha_0) & 0 & 0 & \cos(\alpha_0)\sin(\alpha_0) \\0 & 0 &
    0 & 0 \\0 & 0 & 0 & 0 \\\cos(\alpha_0)\sin(\alpha_0) & 0 & 0 &
    \sin^2(\alpha_0)\end{array}\right).\ee 
In the following we would like to consider the pull-back of the Hermitian
tensor proposed in the previous discussion on a orbit having this state as
fiducial state. We find in this regard by applying on $\ket{0}$ the group
action defined by 
\be Q= \text{SU(2)} \times \text{SU(2)},\ee
the topological distinguished orbits  
\be \mathcal{O}_{\alpha_0}\cong Q/Q_{\ket{0}_{\alpha_0}},\ee
in dependence of the choice $\alpha_0$ in the fiducial state $\ket{0}$.
They admit according to M. Sinolecka et al. \cite{Kus:2002}, the following topologies,
\be \mathcal{O}_{0}\cong S^2\times S^2\ee
\be \mathcal{O}_{0<\alpha_0<\pi/4}\cong S^2 \times (S^3/\Z_2)\ee
\be \mathcal{O}_{\pi/4}\cong S^3/\Z_2\ee
yielding a stratification of the space of pure states into
\be \C P^3 = \bigcup_{\alpha_0\in [0,\pi/4]} \mathcal{O}_{\alpha_0}.\ee
We will show in the following that each of them coincide with one 4-dimensional orbit of separable states ($\alpha_0=0$), one 3-dimensional orbit of maximal entangled states ($\alpha_0=\pi/4$) and the state-space-volume filling foliation into 5-dimensional orbits of intermediate entangled states ($0<\alpha_0<\pi/4$).

  \begin{remark}
Note that the real parameter $\alpha_0$ may be identified with the parameter of a
curve intersecting all orbits in the state space exactly once. The associated
vector field generating this curve turns out to be transversal to the tangent
spaces of the orbits where the curves intersects. In particular one may
associate to such a curve an action of a 1-dimensional subgroup in Aut($\H\otimes
\H$)=U(4), being not a subgroup in SU(2)$\times$ SU(2).
    
  \end{remark}

\begin{figure}[htp]
\centerline{\includegraphics[width=2.9in]{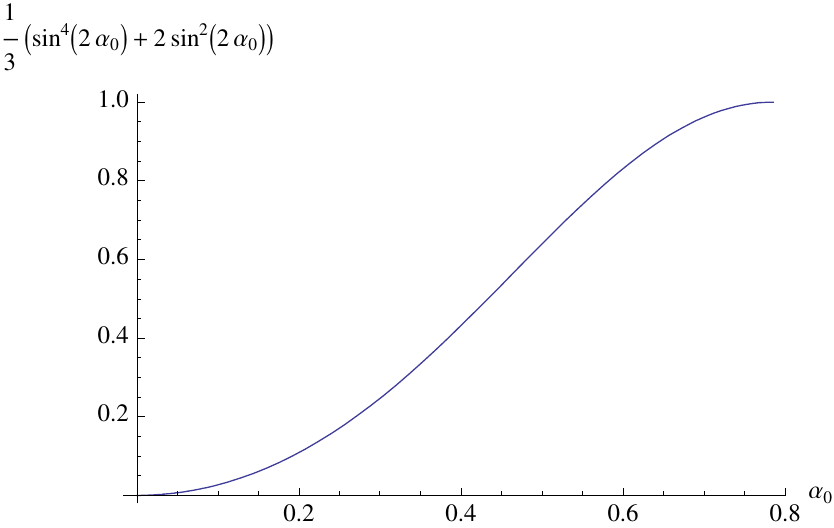}}
\vspace*{8pt}
\caption{Evaluation of $\frac{N^2}{4(N^2-1)}\text{Tr}(C^TC)$ for a Schmidt-decomposed 2-qubit.\label{fig1}}
\end{figure}

By applying the previous discussion, we find \emph{in dependence of the choice
  of the fiducial state}  $\ket{0}\bra{0}$ that the pull-back of the Hermitian
tensor $K$ on the orbit $\mathcal{O}_{\alpha_0}$ admits the following form  
$$ \left(
\begin{array}{llllll}
 1 & i \cos \left(2 \alpha _0\right) & 0 & \sin \left(2 \alpha _0\right)
   & 0 & 0 \\
 -i \cos \left(2 \alpha _0\right) & 1 & 0 & 0 & -\sin \left(2 \alpha
   _0\right) & 0 \\
 0 & 0 & \sin ^2\left(2 \alpha _0\right) & 0 & 0 & \sin ^2\left(2 \alpha
   _0\right) \\
 \sin \left(2 \alpha _0\right) & 0 & 0 & 1 & i \cos \left(2 \alpha
   _0\right) & 0 \\
 0 & -\sin \left(2 \alpha _0\right) & 0 & -i \cos \left(2 \alpha
   _0\right) & 1 & 0 \\
 0 & 0 & \sin ^2\left(2 \alpha _0\right) & 0 & 0 & \sin ^2\left(2 \alpha
   _0\right)
\end{array}
\right),$$
which due to the decomposition  \be K_{jk}=K_{(jk)}+iK_{[jk]} \ee gives the imaginary anti-symmetric part $i(K_{[jk]})$,
$$\left(
\begin{array}{llllll}
 0 & i \cos \left(2 \alpha _0\right) & 0 & 0 & 0 & 0 \\
 -i \cos \left(2 \alpha _0\right) & 0 & 0 & 0 & 0 & 0 \\
 0 & 0 & 0 & 0 & 0 & 0 \\
 0 & 0 & 0 & 0 & i \cos \left(2 \alpha _0\right) & 0 \\
 0 & 0 & 0 & -i \cos \left(2 \alpha _0\right) & 0 & 0 \\
 0 & 0 & 0 & 0 & 0 & 0
\end{array}
\right),$$
and the real symmetric part $(K_{(jk)})$,
$$\left(
\begin{array}{llllll}
 1 & 0 & 0 & \sin \left(2 \alpha _0\right) & 0 & 0 \\
 0 & 1 & 0 & 0 & -\sin \left(2 \alpha _0\right) & 0 \\
 0 & 0 & \sin ^2\left(2 \alpha _0\right) & 0 & 0 & \sin ^2\left(2 \alpha
   _0\right) \\
 \sin \left(2 \alpha _0\right) & 0 & 0 & 1 & 0 & 0 \\
 0 & -\sin \left(2 \alpha _0\right) & 0 & 0 & 1 & 0 \\
 0 & 0 & \sin ^2\left(2 \alpha _0\right) & 0 & 0 & \sin ^2\left(2 \alpha
   _0\right)
\end{array}
\right).$$ 
By furthermore decomposing the latter part, in the block-diagonal matrices $A, B$ and the two equal block-off-diagonal matrices $C$, according to
\be (K_{(jk)})=\left(\begin{array}{cc}A & C \\C & B\end{array}\right),\ee
we identify 
\be A=B=\left(
\begin{array}{lll}
 1 & 0 & 0 \\
 0 & 1 & 0 \\
 0 & 0 & \sin ^2\left(2 \alpha _0\right)
\end{array}
\right),\ee
and
\be C=\left(
\begin{array}{lll}
 \sin \left(2 \alpha _0\right) & 0 & 0 \\
 0 & -\sin \left(2 \alpha _0\right) & 0 \\
 0 & 0 & \sin ^2\left(2 \alpha _0\right)
\end{array}
\right).\ee
By using the block-off-diagonal matrices $C$, we find the  entanglement monotone proposed in \cite{Schlienz:1995} according to
\be \frac{N^2}{4(N^2-1)}\text{Tr}(C^TC)= 
\frac{1}{3} \left(\sin ^4\left(2 \alpha _0\right)+2 \sin ^2\left(2 \alpha_0\right)\right),\label{beta}\ee
whose dependence on the angle $\alpha_0\in [0,\pi/4]$  is illustrated in Fig.\ref{fig1}. 
 The maximum is related, as expected, to a maximal entangled state ($\alpha_0=\pi/4 \approx 0.78$),
\be \ket{0}_{\a_0=\pi/4}=\frac{1}{\sqrt{2}}\begin{pmatrix}
1 \\
0 
\end{pmatrix}\otimes \begin{pmatrix}
1 \\
0
\end{pmatrix} + \frac{1}{\sqrt{2}}\begin{pmatrix}
0 \\
1 
\end{pmatrix}\otimes \begin{pmatrix}
0 \\
1
\end{pmatrix}.\label{phi+}\ee
Here we find, according to Proposition 1(a), a vanishing symplectic tensor \be (K_{[jk]}|_{\a_0=\pi/4})=0,\label{s=0}\ee 
and a symmetric part, which reads after diagonalization
\be (K_{(jk)}|_{\a_0=\pi/4})\mapsto\left(
\begin{array}{llllll}
 2 & 0 & 0 & 0 & 0 & 0 \\
 0 & 2 & 0 & 0 & 0 & 0 \\
 0 & 0 & 2 & 0 & 0 & 0 \\
 0 & 0 & 0 & 0 & 0 & 0 \\
 0 & 0 & 0 & 0 & 0 & 0 \\
 0 & 0 & 0 & 0 & 0 & 0
\end{array}
\right).
\\\ee
This recovers the fact that the orbit \be \mathcal{O}_{\pi/4}\cong S^3/\Z_2\ee
of maximal entangled states is 3-dimensional and Lagrangian \cite{Bengtsson:2007},\cite{Bengtsson:2001yd}.\\
On the other extreme, by considering a separable state
\be \ket{0}_{\a_0=0}=\begin{pmatrix}
1 \\
0 
\end{pmatrix}\otimes \begin{pmatrix}
1 \\
0
\end{pmatrix}\ee
we get according to Proposition 1 (b) a vanishing block-off-diagonal block matrix $C$ and a direct sum 
\be  (K_{jk}|_{\a_0=0})=   \left(
\begin{array}{lll}
 1 & i & 0 \\
 -i & 1 & 0\\
 0 & 0 & 0 \\
 \end{array}
\right)\oplus \left(
\begin{array}{lll}
 1 & i & 0 \\
 -i & 1 & 0\\
 0 & 0 & 0 \\
 \end{array}
\right)\ee
of two decoupled Hermitian tensors each defined on one factor of $S^2\times S^2$, the 4-dimensional orbit of separable states admitting the corresponding pull-back tensor 
\be  (K_{jk}|_{\a_0=0})=\left(
\begin{array}{llllll}
 1 & i & 0 & 0 & 0 & 0 \\
 -i & 1 & 0 & 0 & 0 & 0 \\
 0 & 0 & 0 & 0 & 0 & 0 \\
 0 & 0 & 0 & 1 & i & 0 \\
 0 & 0 & 0 & -i & 1 & 0 \\
 0 & 0 & 0 & 0 & 0 & 0
\end{array}
\right).\ee
Finally we find that the 5-dimensional orbits
\be \mathcal{O}_{\a_0\in(0,\pi/4)}\cong S^2 \times (S^3/\Z_2)\ee
of intermediate entangled states admit each of them a symmetric tensor \be (K_{(jk)}|_{\a_0\in(0,\pi/4)}),\ee which reads after diagonalization 
$$\left(
\begin{array}{llllll}
 0 & 0 & 0 & 0 & 0 & 0 \\
 0 & 1-\sin \left(2 \alpha _0\right) & 0 & 0 & 0 & 0 \\
 0 & 0 & 1-\sin \left(2 \alpha _0\right) & 0 & 0 & 0 \\
 0 & 0 & 0 & 2 \sin ^2\left(2 \alpha _0\right) & 0 & 0 \\
 0 & 0 & 0 & 0 & \sin \left(2 \alpha _0\right)+1 & 0 \\
 0 & 0 & 0 & 0 & 0 & \sin \left(2 \alpha _0\right)+1
\end{array}
\right).$$
In this regard one finds that the concurrence of the fiducial state $\ket{0}_{\alpha_0}$
computed via the square root of the tangle \cite{Coffman:1999jd},
\be \tau :=\text{Det}(\Tr_1(\ket{0}\bra{0}) ),\ee 
\be \sqrt\tau =\frac{1}{2}\sin(2\alpha_0),\ee
is directly related to the eigenvalues of the symmetric part of the pulled back tensor (resp. the Gram matrix used in \cite{Kus:2001}).\\ 
The anti-symmetric part on the other hand, stands in a correspondence to the pull back on separable states ($\a_0=0$) up to a multiplicative factor $\cos(2\a_0)$ due to
\be  (K_{[jk]}|_{\a_0\in(0,\pi/4)})=\cos(2\a_0)(K_{[jk]}|_{\a_0=0}),\ee
which become according to (\ref{s=0}) degenerate for maximal entangled states at $\a_0=\pi/4$. 

  \begin{remark}
We have seen that the embedding via a local unitary action provided a pull-back tensor whose invariants are associated with an entanglement monotones known as the concurrence $\sqrt\tau$. We may close a circle by using the latter for the purpose of identifying an embedding in the following way: One finds here by identifying the fiducial state 
\be \ket{0}_{\a_0}\equiv \cos(\a_0)(1,0,0,0)^T+\sin(\a_0)(0,0,0,1)^T :=(Z_0,Z_1,Z_2,Z_3)^T\ee
with four complex coordinates that 
\be Z_0Z_3-Z_1Z_2 = \sin(2 \alpha_0),  \ee
i.e.
\be Z_0Z_3-Z_1Z_2 =  2\sqrt\tau.\label{P-emb} \ee
The concurrence on the right hand side of the relation (\ref{P-emb}) may therefore associated to a family of embeddings,  
\be \iota:\mathcal{M}_{\a_0}\hookrightarrow \mathcal{P}(\C^2\otimes \C^2)\ee  in the a projective Hilbert space $\mathcal{P}(\C^4)$,  parameterisized by $\a_0\in[0,\pi/4]$, providing a generalization of the case $\a_0=0$,
\be Z_0Z_3-Z_1Z_2 =0,\ee
\be \iota:\C P^1\times \C P^1 \hookrightarrow \mathcal{P}(\C^2\otimes \C^2),\ee
 known as the \emph{Segre embedding}.    
  \end{remark}

\section{Conclusion and Outlook}

In this paper we have considered our previous construction of classical
tensor fields out of quantum states, to investigate what happens when we
deal with pure states of composite bipartite systems. In particular we
have shown that it is possible to extract information for separable states
and entangled states, both from the Riemannian structure and from the
symplectic structure. Our tensor fields allow us to extract information
that within the usual treatment would require dealing with different
orbits of the local unitary transformation group. One may foresee that other
tensorial quantities like the curvature of the metric tensor may furnish
further information on the entanglement. These tensorial descriptions may
be put to use to perform also generic nonlinear transformations. This may
turn out to be rather useful for generic density states which form a
stratified manifold rather than a smooth manifold, in particular for those
strata which are orbits of nonlinear actions of the general linear
group. We shall consider in the near future these questions in connection
with density states and the GNS construction.

\section*{Acknowledgments}

We thank D. D\"urr, A. Kossakowski, and E. C. G. Sudarshan for helpful comments.

\end{document}